# Socio-Economic Consequences of Generative AI: A Review of Methodological Approaches

Carlos J. Costa
Advance/ISEG (Lisbon School of Economics & Management), Universidade de Lisboa, Portugal
cjcosta@iseg.ulisboa

Joao Tiago Aparicio
INESC-ID and Instituto Superior Técnico, Universidade de Lisboa; LNEC, Department of Transport, Av. do Brasil 101, Lisboa, 1700-075, Lisboa, Portugal
joao.aparicio@tecnico.ulisboa.pt

Manuela Aparicio
NOVA Information Management School (NOVA IMS), Universidade Nova de Lisboa, Portugal
manuela.aparicio@novaims.unl.pt

*Abstract* — **The widespread adoption of generative artificial intelligence (AI) has fundamentally transformed technological landscapes and societal structures in recent years. Our objective is to identify the primary methodologies that may be used to help predict the economic and social impacts of generative AI adoption. Through a comprehensive literature review, we uncover a range of methodologies poised to assess the multifaceted impacts of this technological revolution. We explore Agent-Based Simulation (ABS), Econometric Models, Input-Output Analysis, Reinforcement Learning (RL) for Decision-Making Agents, Surveys and Interviews, Scenario Analysis, Policy Analysis, and the Delphi Method. Our findings have allowed us to identify these approaches' main strengths and weaknesses and their adequacy in coping with uncertainty, robustness, and resource requirements.**

## I. INTRODUCTION

In recent years, generative artificial intelligence (AI) usage has revolutionized technological landscapes and profoundly reshaped the societal structure. This transformative force, marked by its ability to generate novel content, ideas, and solutions autonomously, has sparked unprecedented levels of innovation across various sectors [14, 22, 42, 52, 54]. However, alongside its potential for progress, generative AI also brings forth a plethora of uncertainties and complexities, emphasizing the urgent need for structured evaluation methodologies.

As society struggles with the implications of this technological revolution, the imperative to develop robust and scalable analytical frameworks becomes increasingly evident [17, 22]. The unpredictable yet extensive impacts of generative AI underscore the necessity of comprehensively understanding its implications. From its effects on workforce dynamics to its influence on socio-economic structures and cultural norms, the multifaceted nature of AI innovation demands a nuanced evaluation approach.

In the research reported here, we embark on a comprehensive exploration motivated by the tension between the unforeseeable shifts brought about by generative AI and the necessity for structured evaluation methodologies. This exploration is driven by the acknowledgment that while the potential benefits of AI innovation are vast, so are the associated challenges and risks. Therefore, our effort aims to solve the complexities inherent in this technological revolution and provide insights to inform strategic decision-making and policy formulation in an era of unprecedented change.

Our investigation examines various methodologies, ranging from traditional econometric models to cutting-edge reinforcement learning techniques tailored for decision-making agents. Agent-Based Simulation (ABS), Input-Output Analysis, Surveys and Interviews, Scenario Analysis, Policy Analysis, and the Delphi Method all feature prominently, each offering unique insights into the intricate dynamics entwined with the proliferation of generative AI. They are synthesizing insights gleaned from these methodologies. Recognizing the nuanced interplay between technological advancement and its societal ramifications, we argue that such an integrative approach is essential for comprehensively understanding the transformative effects of generative AI.

Moreover, by shedding light on the multifaceted consequences of AI innovation, our research aims to facilitate informed decision-making and policy formulation in the face of unprecedented technological change. Through this integrative lens, we aspire to contribute to a deeper comprehension of the societal implications of AI innovation, empowering stakeholders to navigate the complexities of an increasingly AI-driven world with foresight and discernment.

## II. LITERATURE REVIEW

From the literature [1-19, 49], the following methodologies are commonly used to analyze the impact of adopting new technology. The choice of method depends on the specific research questions, available data, and the complexity of the economic system under consideration [49]. We may identify as possible methods the figure. 1.

Agent-based modeling (ABM) consists of modeling individual entities (agents) and their interactions within a system [1]. Each agent follows predefined rules or behaviors, and the simulation explores the emergent properties resulting from these interactions. ABM is used to understand complex systems and study how local interactions lead to global patterns [2, 51]. ABM allows for modeling complex behavior patterns and provides valuable information about the dynamics of real-world systems

[3]. It can incorporate learning techniques such as neural networks and evolutionary algorithms to enable realistic learning and adaptation [9]. ABM has broad applications in various fields, including ecology, evolution, economics, and social sciences [10].

1. Agent-Based Modeling (ABM).
2. Econometric Models.
3. Input-Output Analysis.
4. Reinforcement Learning (RL).
5. Surveys and Interviews.
6. Scenario Analysis.
7. Policy Analysis.
8. Delphi Method.

Fig.1- Methods

Econometric models use statistical methods to analyze relationships between economic variables [4]. These models are based on historical data and aim to estimate and quantify the impact of different factors on economic outcomes [7]. Econometric models are commonly used in forecasting and policy analysis [11, 50]. Econometric models provide a robust framework for analyzing the impact of generative AI on economics across multiple dimensions. These models enable researchers to examine AI adoption patterns, labor market dynamics, productivity growth, market competition, policy implications, and long-term economic outlook. By leveraging historical data and statistical methods, econometricians can quantify the effects of generative AI on variables such as employment levels, wages, innovation, market dynamics, and economic welfare. Econometric analyses shed light on how AI-driven innovations influence productivity, consumer behavior, firm profitability, and regulatory responses. Through rigorous modeling and simulation, policymakers can anticipate the economic consequences of AI adoption and design policies that foster innovation, mitigate risks, and ensure equitable outcomes in AI-driven economies. In essence, econometric models serve as invaluable tools for understanding and navigating the complex interactions between generative AI and economics, providing insights that inform decision-making and shape the future trajectory of economic development.

Input-output analysis examines the interdependencies between different economic sectors [5, 20]. It quantifies how changes in one sector impact others by analyzing the flows of goods and services [6]. Input-output models help understand the ripple effects of economic changes [6]. Input-output analysis, which examines the interdependencies among economic sectors and quantifies the flow of goods and services, is a valuable tool for predicting the impact of generative AI in economics. By understanding how changes in one sector affect others, input-output models can anticipate the ripple effects of AI-generated goods and services across various industries. For instance, as generative AI technologies become prevalent in creative content production, such as art and music, input-output analysis can forecast the downstream effects on related sectors like entertainment and advertising. Additionally, it can assess the potential labor displacement resulting from increased automation facilitated by generative AI. By simulating different scenarios and analyzing economic activity flows, stakeholders can better anticipate and manage the consequences of AI-driven innovation, making input-output analysis essential for policymakers, businesses, and researchers navigating the complexities of the evolving economic landscape.

Reinforcement learning (RL) is a machine learning paradigm where an agent learns to make decisions by interacting with an environment [13]. In RL for decision-making agents, the focus is on modeling the decision-making processes of individual entities. Agents learn optimal strategies through trial-and-error interactions [13]. RL involves the exploration of solution spaces within a simulated environment to optimize decision-making [14]. RL can be combined with agent-based models (ABMs) to address the self-organizing dynamics of social phenomena and explore the space of possibilities that emerge from considering different types of incentives [15]. Reinforcement learning (RL) presents a potent methodology for forecasting the implications of generative artificial intelligence (AI) within economics. By employing RL, researchers can simulate the decision-making processes of economic agents interacting with generative AI systems. RL algorithms allow for iterative learning through trial-and-error interactions, enabling agents to optimize their strategies based on feedback. Additionally, RL facilitates the exploration of solution spaces within simulated environments, aiding in identifying optimal decision-making strategies amidst complex economic scenarios influenced by generative AI. Integration with agent-based models (ABMs) further enhances understanding by capturing heterogeneous agents' dynamic interactions and emergent behaviors. RL techniques can also inform incentive design and policy formulation, optimizing reward structures and decision mechanisms to guide economic agents' behavior in AI-driven environments. Overall, RL serves as a valuable tool for anticipating and shaping the economic impacts of generative AI, providing insights crucial for designing robust and adaptive economic policies.

Surveys and interviews involve collecting data through structured questionnaires [53, 57] or direct interviews with individuals. In the context of economic analysis, surveys and interviews can gather opinions, preferences, and insights from key stakeholders, such as businesses, policymakers, or consumers [12]. Surveys and interviews serve as invaluable tools in predicting the impact of generative AI in economics by capturing diverse stakeholders' perspectives and insights. These methods facilitate understanding current perceptions and awareness of generative AI technologies, identifying potential use cases, and assessing associated opportunities and challenges. Through gathering expert opinions and stakeholder feedback, surveys and interviews enable forecasting economic impacts, including productivity gains, changes in employment dynamics, and implications for market structures and income distribution. Moreover, the data collected can inform the development of policies and strategic initiatives to maximize benefits and minimize risks associated with integrating generative AI into economic systems, ultimately contributing to informed decision-making and fostering inclusive economic growth.

Scenario analysis involves creating and exploring multiple future scenarios based on different assumptions or conditions. It helps decision-makers understand potential outcomes under various circumstances and develop strategies to navigate uncertainty [16]. Scenario analysis is a strategic forecasting technique that can be employed to anticipate the impact of generative AI in economics. By identifying key drivers such as technological advancements, regulatory frameworks, and societal attitudes, scenario analysis involves creating a variety of plausible future scenarios. These scenarios explore different combinations of factors to understand their potential implications on economic aspects like productivity, employment, and market dynamics. Through quantitative modeling and qualitative assessments, decision-makers can evaluate the risks and opportunities associated with each scenario, informing strategic planning and policy development. Scenario analysis is an iterative process, allowing stakeholders to continuously refine their understanding of the potential impacts of generative AI and adapt their strategies accordingly. Ultimately, scenario analysis provides a structured approach to navigating uncertainty and making informed decisions in the face of emerging technologies like generative AI.

Policy analysis evaluates the effectiveness of policies and interventions in achieving specific goals. In the context of economic impact analysis, policy analysis assesses how different regulatory frameworks, incentives, or interventions may influence economic outcomes [17, 55, 56]. Policy analysis offers a valuable framework for predicting the impact of generative AI in economics by systematically evaluating potential outcomes and crafting appropriate responses. It involves identifying how generative AI could influence economic dynamics, including changes in production processes, labor markets, consumer behavior, and market structures. Through scenario planning and cost-benefit analysis, policymakers can anticipate challenges and opportunities associated with adopting generative AI technologies, informing the design of regulatory frameworks and incentive structures. By engaging diverse stakeholders and considering ethical considerations, policymakers can develop evidence-based policies to maximize the benefits of generative AI while mitigating risks, thus ensuring responsible innovation and fostering economic growth in the AI-driven era.

The Delphi Method is a structured, iterative forecasting technique that gathers expert opinions through surveys or questionnaires [18]. Experts anonymously provide feedback on a specific topic, and the process continues until consensus or stability is reached. The method is useful in situations with high uncertainty [19]. The Delphi Method, a structured forecasting technique, can be employed to predict the impact of generative AI in economics by assembling a diverse panel of experts in economics and artificial intelligence. Through a series of iterative surveys, experts anonymously provide feedback on the potential effects of generative AI in various economic domains, such as productivity, labor markets, and decision-making processes. The process involves defining the study scope, designing open-ended questionnaires, aggregating responses, refining questions based on initial analysis, and iteratively repeating surveys until consensus or stability is achieved. By synthesizing expert opinions, the method generates predictions about the impact of generative AI in economics, considering diverse perspectives and uncertainties. These predictions can inform policymakers, researchers, and industry professionals in making informed decisions and preparations for the future.

By employing these methodologies, researchers can gain valuable insights into the economic implications of generative AI and its impact on various sectors of the economy. A preliminary evaluation (in Table I) may consist of a general identification of each approach's main strengths and weaknesses.

TABLE I. STRENGTH AND WEAKNESSES

| Method | Strengths | Weaknesses | Ref. |
|---|---|---|---|
| Agent-Based Modeling (ABM) | Simulates dynamic interactions, explores emergent impacts | Data intensive relies on model assumptions, complex | [3], [21]. |
| Econometric Models | Predicts economic impacts, leverages historical data | Simplified assumptions, limited to economic aspects, potentially inaccurate | [22]. |
| Input-Output Analysis | Tracks economic flows, predicts production shifts. | Limited to economic impacts, ignores social/cultural aspects, data requirement | [20] |
| Reinforcement Learning (RL) | Adapts to changing environments, predicts system/human interaction evolution. | Extensive training data, limited generalizability, potential biases | [8]. |
| Surveys and Interviews | Gathers direct opinions, understands human perceptions | Limited representativeness, prone to biases, not good for long-term predictions | [12]. |
| Scenario Analysis | Explores different futures, facilitates strategic planning | Relies on scenario quality, doesn't offer precise predictions | [16]. |
| Policy Analysis | Assesses policy impacts, aids decision-making | Focuses on specific policies, overlooks broader changes, requires comprehensive policy options | [17].. |
| Delphi Method | Structured expert opinion gathering, converges on impact range | Subjective, prone to expert bias, not ideal for specific predictions | [18], [19], [23]. |

III. METHODOLOGY

These methods offer diverse approaches to studying and understanding the economic impact of various factors, such as the introduction of new technologies like generative AI. In order to evaluate those approaches, the following criteria must be used: uncertainty, robustness, scalability, and resource requirements. The methodology's alignment with the study's objectives is supported by Johnson and Onwuegbuzie [24], who advocate for mixed methods research as a comprehensive approach to address diverse research questions. Tashakkori and Teddlie further underscore the importance of integrating qualitative and quantitative methods to enhance research depth and breadth [25]. Creswell emphasizes the necessity of selecting a methodology that resonates with the research aims and questions, ensuring ethical and academic integrity [26]. Uncertainty acknowledges the adaptability of methods to unpredictable conditions, while robustness evaluates consistency across diverse scenarios, which is crucial for ensuring reliability [24]. Scalability addresses the applicability of methods to extensive datasets, a key factor in managing computational efficiency [25]. Lastly, resource requirements consider the practicality of method implementation, balancing the need for detailed analysis with the constraints of available resources [26]. This multifaceted evaluation ensures a thorough and grounded selection of methodologies, aligning with both the research objectives and the practical realities of implementation [49]. These references collectively justify the chosen methodological framework, highlighting its capability to yield nuanced insights within a rigorous ethical and theoretical context.

**Uncertainty** indicates a method's ability to acknowledge and handle high levels of uncertainty or unpredictability in the analysis.

1. High: Methods with high uncertainty often involve subjective judgments, assumptions, or scenarios that may vary widely.

2. Moderate: Reflects a balanced consideration of uncertainty, acknowledging potential variations in outcomes or predictions. The method incorporates a degree of flexibility in response to changing conditions or unknowns.

3. Low: Suggests a method that tends to produce more deterministic or predictable outcomes, with limited consideration for uncertain factors. The approach may rely heavily on historical data or well-defined rules.

**Robustness** indicates the resilience of a method to variations or changes in input parameters, assumptions, or conditions.

1. High: A robust method is less sensitive to uncertainties and can produce consistent results across different scenarios.

2. Moderate: Reflects a method's ability to handle variations to some extent. It may be robust under certain conditions but could exhibit sensitivity in others. Results are generally reliable within a defined range of conditions.

3. Low: Suggests that the method may be sensitive to changes in input parameters, assumptions, or conditions. Results may vary significantly under different scenarios.

**Scalability** indicates a method's ability to handle large-scale analyses efficiently.

1. High: Scalable methods can be applied to complex systems or datasets without substantially increasing resource requirements.

2. Moderate: Reflects an intermediate level of scalability. The method is suitable for moderate-sized analyses but may face challenges when applied to larger or more complex scenarios.

3. Low: Suggests limitations in scaling the method to handle large or complex analyses. The approach may become computationally intensive or less practical as the scale of the analysis increases.

**Resource Requirements** imply that the method demands significant resources, including computational power, expertise, time, or data.

1. High: High resource requirements may limit the accessibility or applicability of the method.

2. Moderate: Reflects a balanced level of resource requirements. The method is manageable with standard resources and does not excessively strain computational, human, or data resources.

3. Low: Indicates that the method has minimal resource demands and can be applied with relatively modest computational power, expertise, time, or data.

## IV. RESULTS

The following table summarizes the comparison of different approaches based on uncertainty, robustness, scalability, and resource requirements:

TABLE II. EVALUATING THE APPROACHES

| Approach | Uncertainty | Robustness | Scalability | Resource Requirements |
|---|---|---|---|---|
| Agent-Based Model | High | Moderate | Moderate to High | Moderate to High |
| Econometric Models | Moderate | Moderate | Low to Moderate | Moderate |
| Input-Output Analysis | Moderate | Moderate | Moderate | Low to Moderate |
| Reinforcement Learning | Moderate | Moderate | Low to Moderate | Moderate to High |
| Surveys and Interviews | Moderate | Low to Moderate | Low | Low to Moderate |
| Scenario Analysis | High | Low to Moderate | Low to Moderate | Low to Moderate |
| Policy Analysis | Moderate | Moderate | Low to Moderate | Low to Moderate |
| Delphi Method | High | Low to Moderate | Low to Moderate | Low to Moderate |

These ratings are intended for a general comparison. The appropriateness of each approach depends on the specific context and research objectives, yet the detail aimed with Table II is as follows:

High uncertainty is attributed to ABM due to its reliance on subjective judgments and assumptions about individual agent behaviors and interactions within a system [21]. ABM demonstrates moderate robustness as it can handle variations to some extent, but results may still exhibit sensitivity to changes in initial conditions or agent behaviors [28]. ABM is rated as moderate to high in scalability as it can be applied to complex systems without substantially increasing resource requirements [29]. Moderate to high resource requirements are associated with ABM due to its need for computational power to simulate numerous agents and interactions [30].

Econometric models are assigned moderate uncertainty as they rely on historical data and statistical methods, balancing deterministic outcomes and potential variations [11]. Like uncertainty, econometric models demonstrate moderate robustness, providing reliable results within a defined range of conditions, but may be sensitive to changes in underlying economic factors [8]. Econometric models are rated low to moderate in scalability as they may face challenges when applied to larger or more complex datasets [33]. Moderate resource requirements are associated with econometric models, requiring standard resources for data collection, analysis, and interpretation [34].

Input-output analysis is rated as moderate uncertainty as it relies on historical data and sectoral interdependencies to predict economic impacts, acknowledging potential outcome variations [5]. Input-output analysis demonstrates moderate robustness, providing consistent results within a defined range of conditions but may be sensitive to changes in economic structures or external shocks [36]. Input-output analysis is rated moderate in scalability as it can handle moderate-sized analyses efficiently but may face challenges with larger or more complex datasets [37]. Low to moderate resource requirements are associated with input-output analysis, requiring standard resources for data collection and analysis [38].

Reinforcement learning exhibits moderate uncertainty as it involves iterative learning through trial-and-error interactions, incorporating a balance between deterministic strategies and adaptive behavior [8]. Like uncertainty, reinforcement learning demonstrates moderate robustness, adapting to changing environments while maintaining consistency in decision-making processes [13]. Reinforcement learning is rated low to moderate in scalability as it may face challenges with larger datasets or complex decision-making environments [14]. Moderate to high resource requirements are associated with reinforcement learning, requiring computational power for training data-intensive models [42].

Surveys and interviews demonstrate moderate uncertainty as they gather subjective opinions and insights from stakeholders, acknowledging potential response variations [43]. Surveys and interviews are assigned a low to moderate level of robustness as they may be influenced by biases or variations in respondent perspectives [44]. Surveys and interviews are rated as low in scalability as they may not efficiently handle large-scale data collection or analysis [45]. Low to moderate resource requirements are associated with surveys and interviews, requiring standard resources for data collection and analysis [46].

Scenario analysis is attributed to high uncertainty as it involves exploring multiple future scenarios based on different assumptions or conditions, acknowledging the inherent unpredictability of future events [47]. Scenario analysis demonstrates low to moderate robustness as it relies on the quality of scenarios created and may not offer precise predictions of future outcomes [48]. It is rated low to moderate in scalability as it may face challenges with large-scale scenario creation or analysis [40]. Low to moderate resource requirements are associated with scenario analysis, requiring standard resources for scenario creation and analysis [40].

Policy analysis exhibits moderate uncertainty as it evaluates the effectiveness of policies and interventions in achieving specific goals, acknowledging potential variations in policy outcomes [39]. Policy analysis demonstrates moderate robustness as it provides insights into the impacts of specific policies but may overlook broader changes or unintended consequences [35]. Policy analysis is rated low to moderate in scalability as it may face challenges with large-scale policy evaluations or analysis [32]. Low to moderate resource requirements are associated with policy analysis, requiring standard resources for policy evaluation and analysis [31].

The Delphi Method is attributed to high uncertainty as it relies on expert opinions and iterative feedback, acknowledging the subjective nature of expert judgments [23]. The Delphi Method demonstrates low to moderate robustness as it converges on a range of potential impacts based on expert consensus but may be influenced by expert biases or subjective interpretations [19]. The Delphi Method is rated as low to moderate in scalability as it may face challenges with large-scale expert panel recruitment or data collection [27]. Low to moderate resource requirements are associated with the Delphi Method, requiring standard resources for expert panel coordination and data analysis [18].

## V. Conclusions

In conclusion, tackling the economic implications of generative AI requires a multidimensional analytical approach due to its complex dynamics. Each methodology discussed offers unique advantages and limitations. Agent-based modeling (ABM) excels in simulating individual behaviors and emergent system properties, making it particularly apt for scenarios characterized by high uncertainty and moderate to high resource demands. Econometric Models offer a historical data-driven perspective, entailing moderate uncertainty and resource requirements. The input-output analysis provides insights into sectoral interdependencies with moderate uncertainty and relatively low to moderate resource demands. Reinforcement Learning (RL) integrates machine learning into decision-making processes with moderate uncertainty and resource needs. Surveys and Interviews offer valuable insights into stakeholder perspectives with moderate uncertainty and low to moderate resource requirements. Scenario Analysis effectively navigates uncertainty, although with a high level of uncertainty and low to moderate resource needs. Policy Analysis evaluates interventions with moderate uncertainty and resource demands. The Delphi Method adeptly manages high uncertainty with relatively low to moderate resource requirements. A selection of methodologies, guided by research goals and contextual considerations, ensures a comprehensive understanding of the economic implications of generative AI.

## Acknowledgment

Thank you to the reviewers who gracefully gave valuable and constructive comments. The authors acknowledge financial support from the Fundação para a Ciência and Tecnologia (FCT Portugal) through research grant numbers ADVANCE-CSG UIDB/04521/2020, UI/BD/153587/2022 (PhD Grant) and research grant UIDB/04152/2020—Centro de Investigação em Gestão de Informação (MagIC).